\documentclass[aps,preprint]{revtex4}
\usepackage{amsfonts}
\usepackage{amsmath}
\usepackage{amssymb}
\usepackage{caption}
\usepackage{subfigure}
\usepackage{graphicx}
\usepackage{epstopdf}
\usepackage{float}
\usepackage{ulem}
\usepackage{cancel}
\usepackage{xcolor}
\usepackage{cleveref}
\newcommand{\myref}[1]{Eq. \eqref{#1}}
\newcommand{\figref}[1]{FIG. \ref{#1}}

\begin{document}
\title{Charged torus-like black holes as heat engines}
\author{Hanwen Feng}
\email{fenghanwen@stu.scu.edu.cn}
\author{Yuchen Huang}
\email{huangyuchen@stu.scu.edu.cn}
\author{Wei Hong}
\email{thphysic_weihong@stu.scu.edu.cn}
\author{Jun Tao}
\email{taojun@scu.edu.cn}
\affiliation{Center for Theoretical Physics, College of Physics, Sichuan University,
Chengdu, 610064, PR China}
\begin{abstract}
We investigate the thermodynamical properties of charged torus-like black holes and take it as the working substance to study the heat engines. In the extended phase space, by interpreting the cosmological constant as the thermodynamic pressure, we derive the thermodynamical quantities by the first law of black hole thermodynamics and obtain the equation of state. Then, we calculate the efficiency of the heat engine in Carnot cycle as well as rectangular cycle, and investigate how the efficiency changes with respect to volume. In addition, to avoid a negative temperature, we emphasize that the charge of this black hole can not be arbitrary. Last, we check the calculation accuracy of a benchmark scheme and discuss the upper bound and lower bound for charged torus-like black hole in the scheme.
\end{abstract}
\maketitle

\section{Introduction}

The pioneering work of Hawking and Bekenstein \cite{Bekenstein:1973ur,Bekenstein:1972tm,Hawking:1974sw,Bekenstein:1974ax} stimulates the interest in the study of black hole thermodynamics. Hawking temperature and Bekenstein-Hawking entropy provides a profound insight into the nature of quantum gravity. In the extended framework, the cosmological constant is considered as a dynamical parameter and the mass of an AdS black hole can be interpreted as the enthalpy of the space \cite{Kastor:2009wy, Dolan:2010ha}. Compared with the classical thermodynamic, people have gradually established four laws of black hole thermodynamics. 

The holographic principle is developed from the  Bekenstein-Hawking entropy, while the thermodynamic behaviour of black holes in Anti-de Sitter (AdS) space reveals the strong coupling gauge theory through various proposed dualities \cite{Susskind:1994vu}. Based on this assumption, Johnson proposed a heat engine defined in the extended thermodynamical space, which takes the AdS black holes as the working substance \cite{Johnson:2014yja}. For a negative cosmological constant, the engine cycle corresponds to a process defined on the space of dual field theories. To understand the holographic heat engine, it's rather important to investigate the microscopic structure. Many different theories have been studied to explore the microscopic property of black holes \cite{Strominger:1996sh,Maldacena:1996gb,Horowitz:1996fn,Emparan:2006it,Lunin:2002qf,Dowker:1997vj}. For further investigation, a concept, i.e. the number density $n$ of the virtual black hole molecules, is introduced  to study the behaviours of the microscopic thermodynamic variables \cite{Wei:2015iwa,Wei:2019uqg}. It can be regarded as the order parameter to measure the microscopic degrees of freedom and it's related to the size or radius of the black hole \cite{Li:2020khm}. From this point, we can take the process of a black hole doing work as the changes in $n$, but it remains a conception and still requires more effort to understand the microscopic structure.

The heat engine is defined in the $P-V$ space as a closed path.  At first, we can calculate the efficiency of the holographic heat engines of black holes with vanishing specific heat at constant volume ($C_{V}=0$) in analytical way. Next, Johnson investigated the efficiency of Born-Infeld black hole in the rectangular cycle \cite{Johnson:2015fva} and then obtained a efficiency formula for heat engines in this rectangular cycle \cite{Johnson:2016pfa,Rosso:2018acz}. When the engine is defined as a rectangular cycle expressed with mass and internal energy of the black hole, the calculation for black holes with $C_{V}\neq 0$ is capable as well \cite{Hennigar:2017apu}.  Since various black holes can be the working substance of a heat engine, and the efficiency is a dimensionless quantity, we can compare different black holes' efficiency and investigate their thermodynamical properties further. To avoid the case where one particular heat engine yields advantages for one specific black hole, Chakraborty and Johnson proposed the benchmarking scheme \cite{Chakraborty:2016ssb, Chakraborty:2017weq}, which separates a complicated cycle into rectangular cycles and calculates the efficiency approximately with numerical method. Recently, black holes in massive gravity have been discussed as heat engines in \cite{Hendi:2017bys,Fernando:2018fpq,Yerra:2020bfx}. Then, researchers have studied the thermodynamics and heat engine efficiency of charged accelerating AdS black holes \cite{Zhang:2018hms,Zhang:2018vqs}, nonlinear black holes \cite{Yerra:2018mni, Balart:2019uok} and the general class of accelerating, rotating and charged Plebanski-Demianski black holes \cite{Debnath:2020zdv}. Moreover, Johnson have investigated the de sitter Black holes in \cite{Johnson:2019ayc}. More work on heat engines can be found in \cite{V.:2019ful,Bhamidipati:2016gel,Mo:2018hav,Hu:2018prt,Debnath:2019mzs,Guo:2019rdk,Ahmed:2019yci,Liu:2017baz,EslamPanah:2018ums,MoraisGraca:2018ofn}.

In this paper, we investigate the heat engine efficiency of the torus-like black hole in Carnot cycle, rectangular cycle and the benchmark cycle. The torus-like black hole is a static solution of the Einstein-Maxwell equation, whose event horizon has $S_{1} \times S_{1} \times \mathbb{R}$ topology \cite{Huang:1995zb,Hong:2020zcf,Lemos:1994xp,Lemos:1995cm,Han:2019kjr}. Each surface of this black hole of the spacetime at constant radius has a toroidal topology which is different from that of the asymptotically flat spacetimes.

This paper is organized as follows. In section \eqref{sec: thermodynamics of charged torus-like black-hole}, we investigate the thermodynamic property of a torus-like black hole and derive the expression for thermodynamic quantities. In section \eqref{Heat engine}, we construct a rectangular heat engine cycle, and then take the torus-like black hole as the working substance to study its efficiency with respect to volume. Since the Carnot heat engine is theoretically with the maximum efficiency, we compare it with the rectangular engine to check. In section \eqref{Scheme}, we calculate the efficiency of benchmark cycle and investigate how the upper bound limits the charge. In the end, the conclusion is given in section \eqref{sec:Conclusion}. 

\section{Thermodynamics of charged torus-like black hole}
\label{sec: thermodynamics of charged torus-like black-hole}
We are interested in charged torus-like black hole, the ansatz for metric can be written as \cite{Huang:1995zb},
\begin{eqnarray}
\mathrm{d} s^{2} &=& -f(r) \mathrm{d}t^{2}+\frac{\mathrm{d}r^{2}}{f(r)}+r^{2}(\mathrm{d} \theta^{2}+\mathrm{d} \psi^{2}), \label{line element}
\end{eqnarray}
with
\begin{eqnarray}
f(r) &=& -\frac{\Lambda r^{2}}{3}-\frac{2 M}{\pi r}+\frac{4 Q^{2}}{\pi r^{2}}, \label{metric}
\end{eqnarray}
where $M$ and $Q$ are mass and electric charge of the black hole, $\Lambda$ is the cosmological constant.
The black hole mass can be obtained by $f(r_{+})=0$, where $r_{+}$ is the event horizon radius,
\begin{eqnarray}
M &=& \frac{12 Q^{2} - \pi r_{+}^{4} \Lambda}{6 r_{+}}. \label{mass with Lambda}
\end{eqnarray}
In 4D AdS spacetime, the negative cosmological constant can be considered as thermodynamical pressure with $P = -\displaystyle\frac{\Lambda}{8 \pi}$ \cite{Kastor:2009wy}, so that $M$ can be rewritten as
\begin{eqnarray}
M &=& \frac{6 Q^{2}+4 P \pi^{2} r_{+}^{4}}{3 r_{+}}. \label{mass with pressure}
\end{eqnarray}
The variation of the mass takes on the form
\begin{eqnarray}
\mathrm{d}M = -\frac{2 Q^{2} - 4 P \pi^{2} r_{+}^{4}}{r_{+}^{2}}\mathrm{d}r_{+} + \frac{4 Q}{r_{+}}\mathrm{d}Q + \frac{4 \pi^{2} r_{+}^{3}}{3}\mathrm{d}P. \label{dM}
\end{eqnarray}
\par According to the definition of surface gravity, the Hawking temperature is only related to the black hole metric \cite{Han:2019kjr},
\begin{eqnarray}
T &=& \frac{{f}'(r_{+})}{4 \pi} = \frac{-12 Q^{2}+3 M r_{+}+ 8 P \pi^{2} r_{+}^{4}}{6 \pi^{2} r_{+}^{3}}. \label{Hawking temperature}
\end{eqnarray}
Substituting \myref{mass with pressure} into \myref{Hawking temperature}, the temperature can be written as
\begin{eqnarray}
T = -\frac{Q^{2} - 2 P^{2} r_{+}^{4}}{\pi^{2} r_{+}^{3}}. \label{temperature}
\end{eqnarray}
With the relation between the Bekenstein-Hawking entropy and the surface area of the event horizon, the entropy of this black hole can be expressed as 
\begin{eqnarray}
S = \pi^{2}r_{+}^{2} \label{entropy}.
\end{eqnarray}
The thermodynamic volume and electric potential at the event horizon are given by
\begin{eqnarray}
V = \left(\frac {\partial M}{\partial P}\right)_{S,Q}= \frac{4 \pi^{2} r_{+}^{3}}{3}, \ \  \Phi=\left(\frac {\partial M}{\partial Q}\right)_{P,S}=\frac{4Q}{r_{+}}. \label{quantities}
\end{eqnarray}
In the extended thermodynamics, the mass of an AdS black hole is interpreted to the enthalpy of spacetime \cite{Kastor:2009wy}. And Eq. \eqref{dM}-\eqref{quantities} shows the first law of thermodynamics is satisfied,
\begin{eqnarray}
\mathrm{d}M &= T \mathrm{d}S + \Phi \mathrm{d}Q + V \mathrm{d}P.
\end{eqnarray}

As the entropy and volume are both the function of black holes horizon, the heat capacity at constant volume vanishes,
\begin{eqnarray}
C_{V} &=& T\left(\displaystyle\frac{\partial S}{\partial T}\right)_{V}= 0.
\label{heat capacity}
\end{eqnarray}
The state equation of the black hole can be derived from  \myref{temperature} and \myref{quantities},
\begin{eqnarray}
P &=& \frac{\pi^{2/3}}{3 \sqrt[3] {6}}\left(\frac{4 Q^{2}}{V^{4/3}} + \frac{3 T}{V^{1/3}}\right)
\label{state}
\end{eqnarray}
As the critical point satisfies the condition that the first-order partial derivative with respect to volume of pressure and the second-order partial derivative are 0, we derive these two partial derivatives to find this point.
\begin{eqnarray}
\frac{\partial P}{\partial V} &=  -\displaystyle\frac{\pi ^{2/3} \left(16 Q^2+3 T V\right)}{9 \sqrt[3]{6} V^{7/3}}. \label{derivatives}
\end{eqnarray}
From \myref{derivatives}, we can see that $\displaystyle\frac{\partial P}{\partial V}=0$ isn't feasible as the square term, temperature and volume are always positive. This result indicates the absence of critical point, and correspondingly, this black hole has no second-order phase transition \cite{Wang:2019kxp}
\begin{figure}
	\centering
	\includegraphics[scale=1.0]{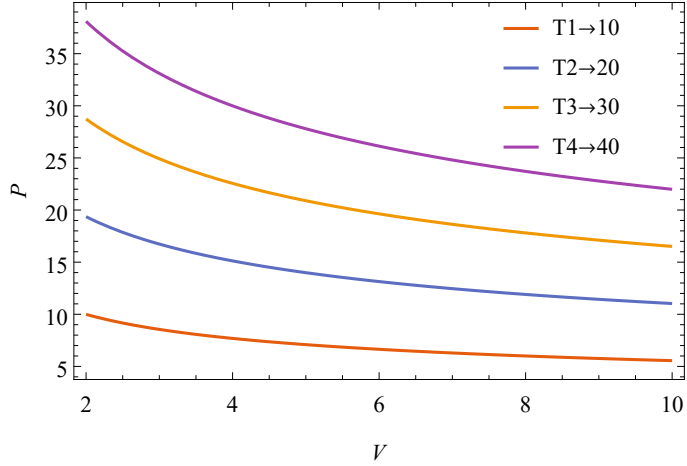}
	\caption{The figure shows $P$ vs $V$ for different temperature. Here $Q$ is fixed at 1.}  
	\label{P vs V}
\end{figure}

For the purpose of studying the behaviour of torus-like black hole as a heat engine, we prefer writing the enthalpy $M$ entirely in terms of $P$ and $V$ by substituting \myref{quantities} into \myref{mass with pressure}, and then the expression yields:
\begin{eqnarray}
M &=& P V + 2 Q^2 \sqrt[3]{\frac{4 \pi^{2}}{3 V}}.
\label{mass with volume}
\end{eqnarray}

It should be noted that, when the torus-like black hole carries no electric charge, i.e. $Q = 0$, its enthalpy $M$ reduces to the same form as that of ideal gas, namely $M = P V$ \cite{Chakraborty:2016ssb}, which will be discussed later.

\section{Charged torus-like black hole as a heat engine}

\label{Heat engine}
\begin{figure}
\centering
\subfigure[Carnot cycle.]{
\begin{minipage}{7cm}
\centering
\includegraphics[scale=0.7]{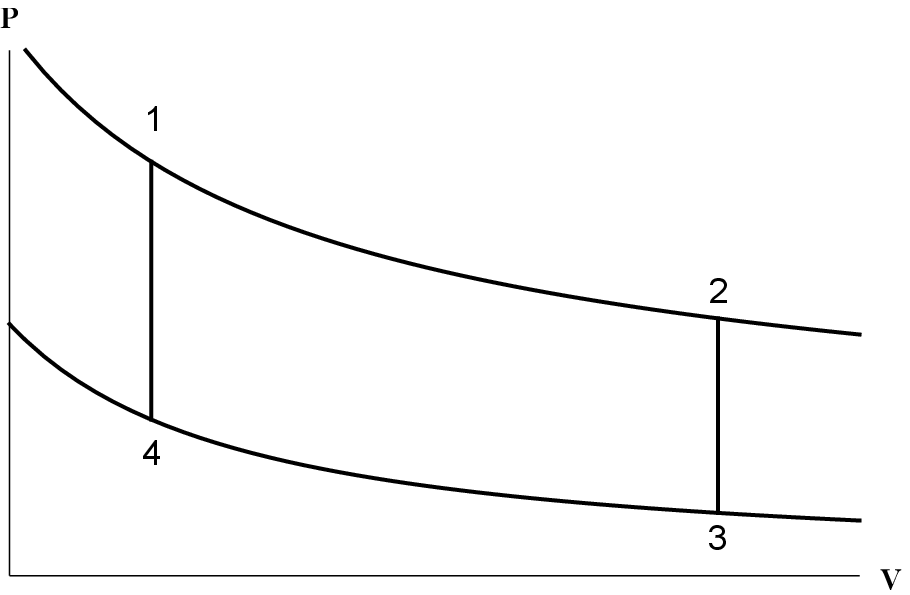}          
\end{minipage}}
\subfigure[Rectangular cycle.]{                    
\begin{minipage}{7cm}
\centering
\includegraphics[scale=0.7]{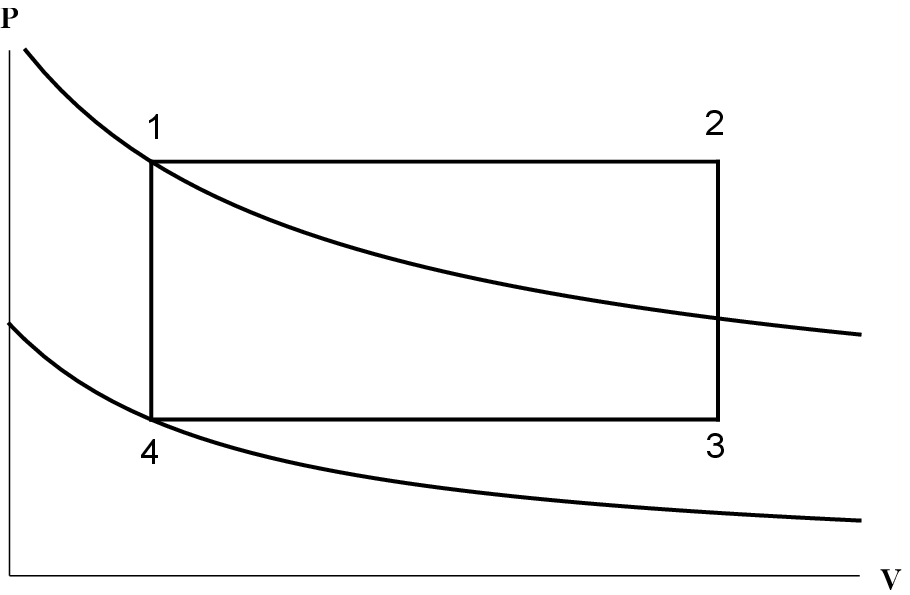}
\end{minipage}}
\caption{The thermodynamical cycles.}                     
\label{Heat engine cycles}
\end{figure}

Since the equation of state for the charged black hole is obtained, we could discuss further by considering this black hole as a heat engine and calculating its efficiency. In the following sections, the black hole is set in thermodynamical cycles and produces work via the $P\mathrm{d}V$ term. We denote the heat absorbed as $Q_{H}$, and the heat exhausted as $Q_{C}$, so that the mechanical work is $W = Q_{H}-Q_{C}$. The efficiency is the ratio of mechanical work $W$ to heat absorbed $Q_{H}$ \cite{Hendi:2017bys},
\begin{eqnarray}
\eta &=& \frac{W}{Q_{H}} = 1 - \frac{Q_{C}}{Q_{H}}. \label{efficiency}
\end{eqnarray} 
The Carnot cycle consists of two isothermal paths and two adiabatic paths and has the highest efficiency. Note that for the black hole engine, the entropy remains constant if the volume doesn't change, so the adiabatic path is equivalent to the isochoric path. Moreover, we define a rectangular cycle with two isochoric and two isobaric paths and compare its efficiency with that of Carnot cycle. We compare the diagram of the two cycles in  \figref{Heat engine cycles}.

The Carnot heat engine is between two different temperatures, we define the higher temperature as $T_{H}$ and the lower one as $T_{L}$, which are connected through the isochoric paths. We can know from \figref{Heat engine cycles} that $T_{H}=T_1=T_2$,  $T_{L}=T_3=T_4$, $V_1=V_4$ and $V_2=V_3$ for Carnot cycle. The heat $Q_{H}$ and $Q_{C}$ can be calculated in the process of isothermal expansion and compression \cite{Hendi:2017bys},
\begin{eqnarray}
Q_{H} &=& T_{H} \Delta S_{1 \rightarrow 2} = T_{H}\frac{\left(3\pi\right)^{2/3}}{2 \times 2^{1/3}}\left(V_{2}^{2/3} - V_{1}^{2/3}\right),\label{Carnot absorb} \\
Q_{C} &=& T_{L} \Delta S_{3 \rightarrow 4} = T_{L}\frac{\left(3\pi\right)^{2/3}}{2 \times 2^{1/3}}\left(V_{3}^{2/3} - V_{4}^{2/3}\right).\label{Carnot discharged}
\end{eqnarray}
Since the volume is connected through isochoric paths, the expression for heat engine efficiency in \myref{efficiency} can be written as:
\begin{eqnarray}
\eta_{c} &=& 1 - \frac{Q_{C}}{Q_{H}} = 1 - \frac{T_{L}}{T_{H}}. \label{Carnot efficiency}
\end{eqnarray}
In order to investigate the relationship between efficiency and volume, one can substitute \myref{mass with Lambda} and \myref{volume} into \myref{Hawking temperature}, the temperature could be rewritten as,
\begin{eqnarray}
T(P,V,Q) &=& \frac{3 \sqrt[3]{6} P V^{4/3}-4 \pi ^{2/3} Q^2}{3 \pi ^{2/3} V}. \label{retemperature}
\end{eqnarray}
Then we can obtain the efficiency of Carnot engine,
\begin{eqnarray}
\eta_{c} &=& 1 - \frac{V_{2} \left(3 \sqrt[3]{6} P_{4} V_{4}^{4/3}-4 \pi ^{2/3} Q^2\right)}{V_{4} \left(3 \sqrt[3]{6} P_{2} V_{2}^{4/3}-4 \pi ^{2/3} Q^2\right)}. \label{carnot efficiency }
\end{eqnarray}
\begin{figure}
	\centering
	\includegraphics[scale=1.0]{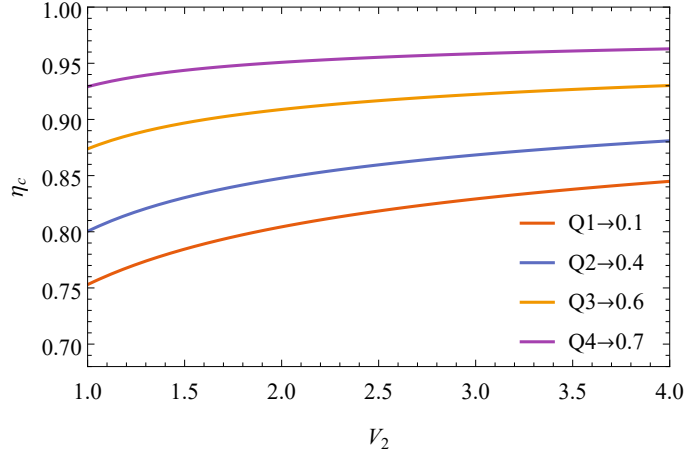}
	\caption{The heat engine efficiency of Carnot cycle versus the black hole volume $V_{2}$ for different charge $Q$, where we set $V_{4}=1$, $P_{2}=4$, and $P_{4}=1$.}
	\label{etac vs V}
\end{figure}

Using the above formula to compute Carnot heat engine efficiency with the volume of this black hole is straightforward. Here we choose parameters  $P_{2}=4$, $P_{4}=1$ and $V_{2}=4$, $V_{4}=1$, with different charge $Q$. Then we can plot the efficiency of Carnot cycle versus the black hole volume $V_{2}$ in \figref{etac vs V}. It shows the efficiency increases monotonously and flattens out with the increase of volume. Note that for a larger volume, the corresponding efficiency is higher.

\begin{figure}
	\centering
	\includegraphics[scale=1.0]{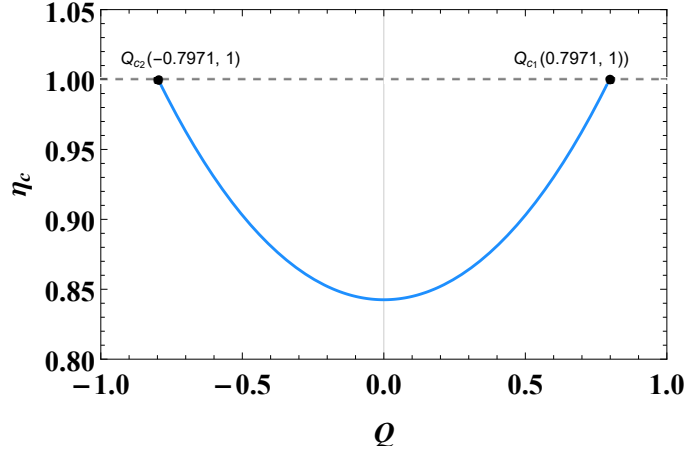}
	\caption{The Carnot heat engine efficiency with varying charge $Q$.}
	\label{etac vs Q}
\end{figure}

It should be noted that the temperature can't be negative, otherwise it makes no sense in physics, so that the charge $Q$ isn't arbitrary. Considering the extremal black hole situation $T = 0$, the maximum charge reads,
\begin{eqnarray}
Q^2  \le \frac{3 \sqrt[3]{6} P V^{4/3}}{4 \pi^{2/3} }.
\end{eqnarray}	
We can set $P_{1}=P_{2}=4$, $P_{3}=P_{4}=1$, $V_{1}=V_{4}=1$, and $V_{2}=V_{3}=4$, after taking both the lower temperature and the higher one into account, the charge of black hole is restricted as $-0.797 \lesssim Q \lesssim 0.797$ to make the expression physical. We can plot the efficiency changing with the varying charge in \figref{etac vs Q} . There are two cut-off points $Q_{c_{1}}= 0.797$ and $Q_{c_{2}}=-0.797$ in the curve, when charge $Q$ reaches the limitation, the black hole temperature $T = 0$ and the heat engine efficiency $\eta = 1$. The cut-off point here reveals the information of Hawking temperature and corresponds to the extremal black holes. Besides, it can be seen from the  \figref{etac vs Q} that the efficiency of the Carnot heat engine has a minimum value,  if the torus-like black hole carries no electric charge.

As we will see, the rectangular cycle is the most natural cycle to consider for all AdS black holes, because it can be generalized to an algorithm, which allows more complex cycles to be numerically calculated. The rectangular cycle goes through isobaric and isochoric paths, although there are four states in the cycle, we can set $P_1=P_2$, $P_3=P_4$, $V_1=V_4$, $V_2=V_3$ as shown in \figref{Heat engine cycles} . The heat engine efficiency for the rectangle cycle can be calculated by a formula deduced in \cite{Johnson:2016pfa, Rosso:2018acz},
\begin{eqnarray}
\eta = 1 - \frac{M_{3}-M_{4}}{M_{2}-M_{1}} =\frac{3 \left(P_2-P_4\right)}{3 P_2-{2 (6 \pi )^{2/3} Q^2}/{\alpha}}, \label{rectangle efficiency}
\end{eqnarray}
where
\begin{eqnarray}
\alpha =V_4 V_2^{1/3}+V_4^{2/3} V_2^{2/3}+V_4^{1/3} V_2. 
\end{eqnarray}

\begin{figure}
	\centering
	\includegraphics[scale=1.0]{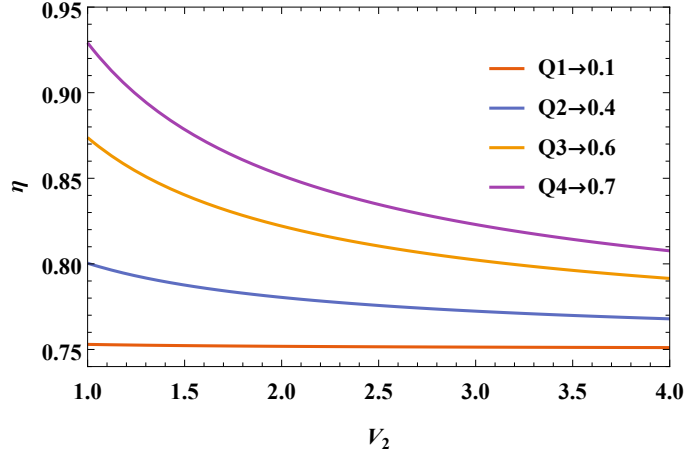}
	\caption{The engine efficiency of rectangle cycle versus the black hole volume $V_{2}$, with parameters set same as the Carnot cycle.}
	\label{eta vs V}
\end{figure}

According to \myref{rectangle efficiency}, we can choose parameters  $P_{1}=P_{2}=4$, $P_{4}=1$ and $V_{1}=V_{4}=1$, and then we plot the heat engine efficiency of rectangle cycle with respect to the volume $V_{2}$ with different charge $Q$  in \figref{eta vs V}. It's obvious that when the charge is fixed, the efficiency curve flattens out as the volume increases and tends to a stable value. Meanwhile, the larger volume $V_{2}$ leads to a lower efficiency, which differs from the situation for the Carnot cycle we mentioned above. On the other hand, for the same volume, the larger $Q$ results in higher efficiency both for Carnot cycle and rectangular cycle.

\section{The benchmark cycle for charged torus-like black hole}\label{Scheme}
In this section, we introduce a benchmark cycle \cite{Chakraborty:2016ssb}, which could be parameterised as a circle. The scheme is adequately complicated as the thermodynamic variables changes on every segment of the cycle, and thus it's more general than regular cycles. This research method can be applied to compare the heat engine efficiency of different black holes. In addition, an upper bound is obtained in Ref. \cite{Hennigar:2017apu} which indicates the efficiency of this cycle is always lower than a specific value. 

Generally, the heat engine efficiency is calculated by numerical method. 
To simplify our calculation, we choose a circular cycle with center of which set as $(P_{o},V_{o})$ and radius as $R$. We overlay our circular cycle onto the $N \times N$ regular lattice of squares, and we require $N$ to be even for simplicity. The side length of every square is $2L/N$, which ensures those squares can cover the circle.  Next we check the cases where two squares share a common isobar and it intersects the circle. Then we calculate the $\Delta M$ of the  left and right endpoints of the isobar.

\begin{figure}
\centering
\includegraphics[scale=1]{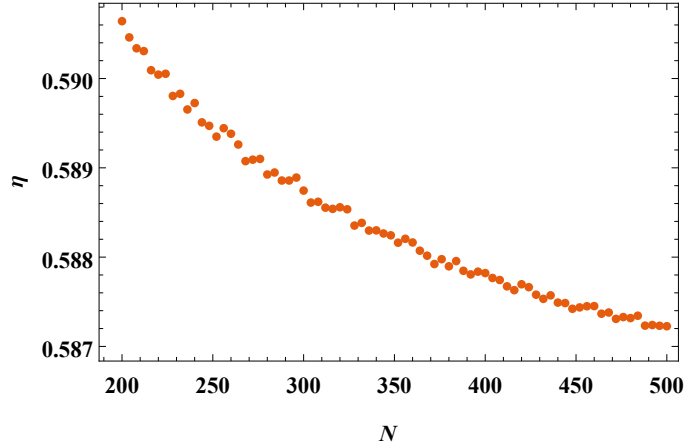}
\caption{Efficiency of the charged torus-like black hole calculated with the numerical method. Here we set the circle origin at $V_{o} = 110$ and $P_{o} = 20$, the charge $Q = 20.0$ and radius $R = 10.0$. The efficiency equals to 0.587226 when $N = 500$.}
\label{Benchmark-N}
\end{figure}

The heat input $Q_{H}$ will be the sum of $\Delta M$ in the upper semicircle and the heat output $Q_C$ is given by the sum of $\Delta M$ in the lower semicircle. Thus we can calculate the efficiency with the numerical method \cite{Chakraborty:2016ssb},
\begin{eqnarray}
\eta &= 1-\displaystyle\frac{Q_{C}}{Q_{H}},
\end{eqnarray}
where,
\begin{eqnarray}
\begin{aligned}
Q_{H} &= \sum\limits_{\textit{i}}(M_{2}^{i}-M_{1}^{i}),\\
Q_{C} &= \sum\limits_{\textit{j}}(M_{3}^{j}-M_{4}^{j}).
\label{numerical}
\end{aligned}
\end{eqnarray}
The result converges to the right value with a larger $N$, as it makes the square smaller and thus fit to the circle more perfectly. \figref{Benchmark-N} shows that as $N$ increases, the efficiency approaches to an exact result. 
\par In a particular case where the specific heat capacity at constant volume $C_{V}=0$, we can calculate the efficiency in an analytical way. Then we can check the the accuracy of the above numerical method by comparing it with an analytical method. The heat engine efficiency  would be \cite{Hennigar:2017apu}
\begin{eqnarray}
\eta &=& \frac{2\pi}{\pi + \displaystyle\frac{2 \Delta M}{R^{2}}}\label{analytical},
\end{eqnarray}
where $\Delta M$ is the enthalpy difference of two points at the left and right ends of the circle,
\begin{eqnarray}
\Delta M &=& M(V_{o}+R, P_{o})-M(V_{o}-R, P_{o})\label{deltaM}.
\end{eqnarray} 
By substituting those parameters of \figref{Benchmark-N} into \myref{analytical}, we obtain the analytical result $\eta_{a} = 0.589322$. Comparing with the results of numerical method when $N$ reaches 500, $\eta_{n} = 0.587226$, we can conclude that  the error of the numerical method is within the acceptable range and the scheme can be applied to the case where analytical method doesn't work.

In the benchmark scheme, an upper bound for the efficiency of black holes with $C_{V}=0$ in circular cycles with the narrow cycle limit and low temperature limit is obtained \cite{Hennigar:2017apu},
\begin{eqnarray}
\eta &\le& \frac{2\pi}{\pi + 4}.
\label{upperbound}
\end{eqnarray}
It's independent of both theory and spacetime dimension, with equality obtained for extremal black holes in the small cycle limit. 

As we mentioned earlier, when the black hole charge $Q=0$, its enthalpy $M$ is the same as that of ideal gas \cite{Chakraborty:2016ssb}, and we can get the lower limit of the efficiency of the torus-like black hole heat engine. The efficiency of ideal gas is independent of $V$ and spacetime and only depends on the pressure at the center of the cycle and the radius of the circle, as mentioned in \cite{Johnson:2015fva, Chakraborty:2016ssb, Chakraborty:2017weq}. 
\begin{eqnarray}
\eta &=& \frac{2 \pi}{\pi + 4P/R}.
\end{eqnarray}

\begin{figure}
\centering
\includegraphics[scale=1.0]{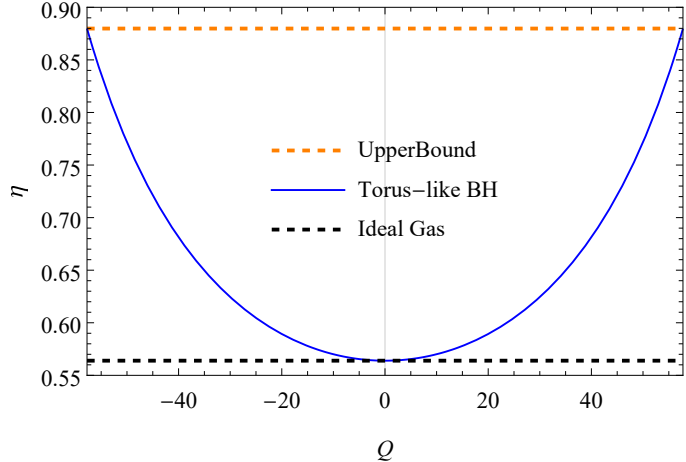}
\caption{This figure plots the efficiency of two black holes with respect to $Q$ in the benchmark cycle. }
\label{Charge-limit}
\end{figure}

We conclude the consideration of benchmarking of the black hole heat engines, and present it in \figref{Charge-limit}. The efficiency of ideal gas model is 0.5639 at $P = 20.0$ and $R =10.0$. When $Q = 57.7431$, $\eta$ reaches the extremal limit 0.8798. As \myref{mass with volume} shows, the mass of torus-like black hole has the same form as the ideal gas when the black hole carries no charge, so the efficiency curve of the black hole and the ideal gas start at the exact point. On the other hand, the top horizontal line is the extremal limit, and it forbids the efficiency to exceed itself.  

\section{Conclusion}\label{sec:Conclusion}
In this paper, we have studied the thermodynamical behaviour of charged torus-like black hole in the extended phase space. By considering the cosmological constant in AdS space as the thermodynamical pressure and the mass as the enthalpy in the first law of thermodynamics, we can obtain all the thermodynamical quantities and the relationship between them.  We derive the equation of state with thermodynamical quantities and find there doesn't exist a critical point in the phase diagram which represents the absence of phase transition. 

Then, we considered charged torus-like black hole as a working substance and studied the holographic heat engine by Carnot cycle and rectangular cycle. The Carnot cycle consists of two isothermal paths and two adiabatic paths, it's always theoretically highest according to the second law of thermodynamics, and thus it provides an upper bound for us to check the calculation. The rectangular cycle is made up of two isochoric paths and two isobaric paths. It works for all black holes, not requiring the specific heat capacity $C_{V}=0$, and it could be an operation unit in other cycles and enable us to calculate the efficiency of the benchmark cycle with numerical methods. From \figref{eta vs V} and \figref{etac vs V}, we conclude that the rectangular efficiency decreases monotonously with respect to the difference between the volume at initial and final states, while the Carnot efficiency does the opposite.

The heat engine cycle in the benchmark scheme is more complicated, and we differentiate the curve into line segments and calculate the efficiency approximately. We compare the numerical result with the analytical result and check the calculation accuracy of numerical methods, then we find the error is acceptable. Due to the fact that the torus-like black hole has the same enthalpy with the ideal gas when $Q = 0$ and that $Q$ increases the efficiency, the efficiency of ideal gas is a lower bound for the torus-like black hole.  On the other hand, the efficiency of a certain class of asymptotically AdS black holes with $C_{V}$ in the circular can't exceed a upper bound as well.

\begin{acknowledgments}
We are grateful to thank Peng Wang and Feiyu Yao for useful discussions. This work is supported by NSFC (Grant No.11947408).

\end{acknowledgments}


\begin{thebibliography}{99}
\bibitem{Bekenstein:1973ur}
J.~D.~Bekenstein,
Black holes and entropy,
Phys. Rev. D \textbf{7} (1973), 2333-2346.

\bibitem{Bekenstein:1972tm}
J.~D.~Bekenstein,
Black holes and the second law,
Lett. Nuovo Cim. \textbf{4} (1972), 737-740.

\bibitem{Hawking:1974sw}
S.~W.~Hawking,
Particle Creation by Black Holes,
Commun. Math. Phys. \textbf{43} (1975), 199-220.

\bibitem{Bekenstein:1974ax}
J.~D.~Bekenstein,
Generalized second law of thermodynamics in black hole physics,
Phys. Rev. D \textbf{9} (1974), 3292-3300.

\bibitem{Kastor:2009wy}
D.~Kastor, S.~Ray and J.~Traschen,
Enthalpy and the Mechanics of AdS Black Holes,
Class. Quant. Grav. \textbf{26} (2009), 195011.

\bibitem{Dolan:2010ha}
B.~P.~Dolan,
The cosmological constant and the black hole equation of state,
Class. Quant. Grav. \textbf{28} (2011), 125020.

\bibitem{Susskind:1994vu}
L.~Susskind,
The World as a hologram,
J. Math. Phys. \textbf{36} (1995), 6377-6396.

\bibitem{Johnson:2014yja}
C.~V.~Johnson,
Holographic Heat Engines,
Class. Quant. Grav. \textbf{31} (2014), 205002.

\bibitem{Strominger:1996sh}
A.~Strominger and C.~Vafa,
Microscopic origin of the Bekenstein-Hawking entropy,
Phys. Lett. B \textbf{379} (1996), 99-104.

\bibitem{Maldacena:1996gb}
J.~M.~Maldacena and A.~Strominger,
Statistical entropy of four-dimensional extremal black holes,
Phys. Rev. Lett. \textbf{77} (1996), 428-429.

\bibitem{Horowitz:1996fn}
G.~T.~Horowitz and A.~Strominger,
Counting states of near extremal black holes,
Phys. Rev. Lett. \textbf{77} (1996), 2368-2371.

\bibitem{Emparan:2006it}
R.~Emparan and G.~T.~Horowitz,
Microstates of a Neutral Black Hole in M Theory,
Phys. Rev. Lett. \textbf{97} (2006), 141601.

\bibitem{Lunin:2002qf}
O.~Lunin and S.~D.~Mathur,
Statistical interpretation of Bekenstein entropy for systems with a stretched horizon,
Phys. Rev. Lett. \textbf{88} (2002), 211303.

\bibitem{Dowker:1997vj}
F.~Dowker, D.~Kastor and J.~H.~Traschen,
U duality, D-branes and black hole emission rates: Agreements and disagreements,
Phys. Rev. D \textbf{58} (1998), 124025.

\bibitem{Wei:2015iwa}
S.~W.~Wei and Y.~X.~Liu,
Insight into the Microscopic Structure of an AdS Black Hole from a Thermodynamical Phase Transition,
Phys. Rev. Lett. \textbf{115} (2015) no.11, 111302.

\bibitem{Wei:2019uqg}
S.~W.~Wei, Y.~X.~Liu and R.~B.~Mann,
Repulsive Interactions and Universal Properties of Charged Anti\textendash{}de Sitter Black Hole Microstructures,
Phys. Rev. Lett. \textbf{123} (2019) no.7, 071103.

\bibitem{Li:2020khm}
R.~Li and J.~Wang,
Thermodynamics and kinetics of Hawking-Page phase transition,
Phys. Rev. D \textbf{102} (2020) no.2, 024085.

\bibitem{Johnson:2015fva}
C.~V.~Johnson,
Born\textendash{}Infeld AdS black holes as heat engines,
Class. Quant. Grav. \textbf{33} (2016) no.13, 135001.

\bibitem{Johnson:2016pfa}
C.~V.~Johnson,
An Exact Efficiency Formula for Holographic Heat Engines,
Entropy \textbf{18} (2016), 120.

\bibitem{Rosso:2018acz}
F.~Rosso,
Holographic heat engines and static black holes: a general efficiency formula,
Int. J. Mod. Phys. D \textbf{28} (2018) no.02, 02.

\bibitem{Hennigar:2017apu}
R.~A.~Hennigar, F.~McCarthy, A.~Ballon and R.~B.~Mann,
Holographic heat engines: general considerations and rotating black holes,
Class. Quant. Grav. \textbf{34} (2017) no.17, 175005.

\bibitem{Chakraborty:2016ssb}
A.~Chakraborty and C.~V.~Johnson,
Benchmarking black hole heat engines, I,
Int. J. Mod. Phys. D \textbf{27} (2018) no.16, 1950012.

\bibitem{Chakraborty:2017weq}
A.~Chakraborty and C.~V.~Johnson,
Benchmarking Black Hole Heat Engines, II,
Int. J. Mod. Phys. D \textbf{27} (2018) no.16, 1950006.

\bibitem{Hendi:2017bys}
S.~H.~Hendi, B.~Eslam Panah, S.~Panahiyan, H.~Liu and X.~H.~Meng, Black holes in massive gravity as heat engines,
Phys. Lett. B \textbf{781} (2018), 40-47.

\bibitem{Fernando:2018fpq}
S.~Fernando,
Massive gravity with Lorentz symmetry breaking: black holes as heat engines,
APS Physics A \textbf{33} (2018), 1850177.

\bibitem{Yerra:2020bfx}
P.~K.~Yerra and C.~Bhamidipati,
Critical Heat Engines in Massive Gravity,
Class. Quant. Grav. \textbf{37} (2020) no.20, 20.

\bibitem{Zhang:2018hms}
J.~Zhang, Y.~Li and H.~Yu,
Thermodynamics of charged accelerating AdS black holes and holographic heat engines,
JHEP \textbf{02} (2019), 144.

\bibitem{Zhang:2018vqs}
J.~Zhang, Y.~Li and H.~Yu, Accelerating AdS black holes as the holographic heat engines in a benchmarking scheme,
Eur. Phys. J. C \textbf{78} (2018) no.8, 645.

\bibitem{Ghaffarnejad:2018gtj}
H.~Ghaffarnejad, E.~Yaraie, M.~Farsam and K.~Bamba,
Hairy black holes and holographic heat engine,
Nucl. Phys. B \textbf{952} (2020), 114941.

\bibitem{Yerra:2018mni}
P.~K.~Yerra and B.~Chandrasekhar,
Heat engines at criticality for nonlinearly charged black holes,
Mod. Phys. Lett. A \textbf{34} (2019) no.27, 1950216.

\bibitem{Balart:2019uok}
L.~Balart and S.~Fernando,
Non-linear black holes in 2+1 dimensions as heat engines,
Phys. Lett. B \textbf{795} (2019), 638-643.

\bibitem{Johnson:2019ayc}
C.~V.~Johnson,
de Sitter Black Holes, Schottky Peaks, and Continuous Heat Engines,
arXiv:1907.05883 [hep-th].

\bibitem{Debnath:2020zdv}
U.~Debnath,
The General Class of Accelerating, Rotating and Charged Plebanski-Demianski Black Holes as Heat Engine,
[arXiv:2006.02920 [gr-qc]].

\bibitem{V.:2019ful}
K.~V.~Rajani, C.~L.~Ahmed Rizwan, A.~Naveena Kumara, D.~Vaid and K.~M.~Ajith,
Regular Bardeen AdS black hole as a heat engine,
Nucl. Phys. B \textbf{960} (2020), 115166.

\bibitem{Bhamidipati:2016gel}
B.~Chandrasekhar and P.~K.~Yerra,
Heat engines for dilatonic Born\textendash{}Infeld black holes,
Eur. Phys. J. C \textbf{77} (2017) no.8, 534.

\bibitem{Mo:2018hav}
J.~X.~Mo and S.~Q.~Lan,
Phase transition and heat engine efficiency of phantom AdS black holes,
Eur. Phys. J. C \textbf{78} (2018) no.8, 666.

\bibitem{Hu:2018prt}
S.~Q.~Hu and X.~M.~Kuang,
Holographic heat engine in Horndeski model with the $k$-essence sector,
Sci. China Phys. Mech. Astron. \textbf{62} (2019) no.6, 60411.

\bibitem{Debnath:2019mzs}
U.~Debnath,
Thermodynamic Black Hole with Modified Chaplygin Gas as a Heat Engine,
Eur. Phys. J. Plus \textbf{135} (2020) no.6, 424.

\bibitem{Guo:2019rdk}
S.~Guo, Q.~Q.~Jiang and J.~Pu,
Heat engine efficiency of the Hayward-AdS black hole,
arXiv:1908.01712 [gr-qc].

\bibitem{Ahmed:2019yci}
W.~Ahmed, H.~Z.~Chen, E.~Gesteau, R.~Gregory and A.~Scoins,
Conical Holographic Heat Engines,
Class. Quant. Grav. \textbf{36} (2019) no.21, 214001.

\bibitem{Liu:2017baz}
H.~Liu and X.~H.~Meng,
Effects of dark energy on the efficiency of charged AdS black holes as heat engines,
Eur. Phys. J. C \textbf{77} (2017) no.8, 556.

\bibitem{EslamPanah:2018ums}
B.~Eslam Panah,
Effects of energy dependent spacetime on geometrical thermodynamics and heat engine of black holes: gravity's rainbow,
Phys. Lett. B \textbf{787} (2018), 45-55.

\bibitem{MoraisGraca:2018ofn}
J.~P.~Morais Gra\c{c}a, I.~P.~Lobo, V.~B.~Bezerra and H.~Moradpour,
Effects of a string cloud on the criticality and efficiency of AdS black holes as heat engines,
Eur. Phys. J. C \textbf{78} (2018) no.10, 823.

\bibitem{Huang:1995zb}
C.~G.~Huang and C.~B.~Liang,
A Torus like black hole,
Phys. Lett. A \textbf{201} (1995), 27-32.

\bibitem{Hong:2020zcf}
W.~Hong, B.~Mu and J.~Tao, Testing the weak cosmic censorship conjecture in torus-like black hole under charged scalar field, Int. J. Mod. Phys. D \textbf{29} (2020) no.12, 2050078.

\bibitem{Lemos:1994xp}
J.~P.~Lemos, Cylindrical black hole in general relativity, Phys. Lett. B {\bf 353} (1995), 46.

\bibitem{Lemos:1995cm}
J.~P.~Lemos and V.~T.~Zanchin, Rotating charged black string and three-dimensional black holes, Phys. Rev. D {\bf 54} (1996), 3840.

\bibitem{Han:2019kjr}
Y.~W.~Han, X.~X.~Zeng and Y.~Hong,
Thermodynamics and weak cosmic censorship conjecture of the torus-like black hole,
Eur. Phys. J. C \textbf{79} (2019) no.3, 252.

\bibitem{Wang:2019kxp}
P.~Wang, H.~Wu and H.~Yang,
Thermodynamics and Phase Transition of a Nonlinear Electrodynamics Black Hole in a Cavity,
JHEP \textbf{07} (2019), 002.



\end{thebibliography}
\end{document}